# A q-deformed logistic map and its implications

Subhashish Banerjee[1,*] and R. Parthasarathy[1,†]

[1]*Chennai Mathematical Institute, Padur PO, Siruseri 603103, India*

A new q-deformed logistic map is proposed and it is found to have concavity in parts of the $x$-space. Its one-cycle and two-cycle non-trivial fixed points are obtained which are found to be qualitatively and quantitatively different from those of the usual logistic map. The stabilty of the proposed q-logistic map is studied using Lyapunov exponent and with a change in the value of the deformation parameter $q$, one is able to go from the chaotic to regular dynamical regime. The implications of this q-logistic map on Parrondo's paradox are examined.

PACS numbers: 05.45.-a,05.45.Ac,02.20.Uw

## I. INTRODUCTION

q-deformation of many classical Lie groups has attracted much attention. It was originally developed in connection with the quantum inverse scattering method [1, 2], and has stimulated much activity in the pursuit of understanding its physical meaning. q-deformed version of Bose commutation relations [3, 4] and non-trivial Fermion anticommutation relations [5] have been used to construct $SU(2)_q$ algebra, Bargmann representation of q-coherent states [6–9] and q-deformed statistics [10, 11]. q-Stirling numbers were introduced [12–14] and studied [15–17]. It has been realized that q-deformation is not merely a mathematical construct but effectively takes into account the interactions in physical systems [18–21]. The q-deformation is non-trivial in the sense that the emerging deformed algebra is no longer linear.

It is therefore natural to use q-deformations suitably in the study of non-linear systems. Discrete dynamical maps in non-linear systems are studied using logistic map [22]. A one-dimensional logistic map is a non-linear difference equation

$$x_{n+1} = A x_n (1 - x_n), \qquad (1)$$

where $A$ is a constant and $x_n$ denotes the value of $x$ after $n$ iterations. Eq. (1) arises, for e.g., in the case of population growth $\frac{dP(t)}{dt} = r(t)P(t)$, where $P(t)$ is the population at a time $t$ and $r(t)$ is the difference between birth and death rates per head of the population. Taking $r(t) = r_0(1 - P(t)/K)$, where $r_0$ and $K$ are constants, and writing $\frac{dP(t)}{dt}$ as the difference between populations at unit intervals of time, one obtains $P_{n+1} = P_n + r_0 P_n(1 - \frac{P_n}{K})$ which gives a nice prediction of the population growth of bacteria, yeast, etc. Setting $x_n = \left(\frac{r_0}{1+r_0}\right)\frac{P_n}{K}$, this equation reduces to Eq. (1) with $A = (1 + r_0)$. In another work [23], the normal logistic and exponential map were used to study the transition from chaotic to regular dynamics induced by stochastic driving.

A q-deformation of the logistic map (1) has been proposed recently [24] in which the map

$$x_{n+1} = A[x_n](1 - [x_n]), \qquad (2)$$

was considered. Here $[x_n] = \frac{x}{1+\epsilon(1-x)}$, and $-1 < \epsilon < \infty$ for $x$ in the interval $[0, 1]$. An important difference between (1) and (2) is that the deformed map (2) is concave in parts of $x$-space while the undeformed map (1) is always convex. Further, the use of (2) showed the rare phenomena of the co-existence of attractors, i.e., the co-existence of normal and chaotic behavior. A comparative study of the co-existing attractors using the Gaussian map and its q-deformed version has been made [25].

The plan of this paper is as follows. In Section II, we introduce our deformed logistic map and motivate its physical implications while in Section III, we study the one and two-cycle fixed points of the map. This is followed, in Section IV, by a discussion of the Lyapunov exponent of the map for various values of the deformation parameter. In Section V, we make a discussion of the phenomena of Parrondo's paradox in the context of logistic maps and make a comparative study of the phenomena on the normal as well as the deformed map. Finally, in Section VI we summarize our results.

---

*Electronic address: subhashish@cmi.ac.in
†Electronic address: sarathy@cmi.ac.in

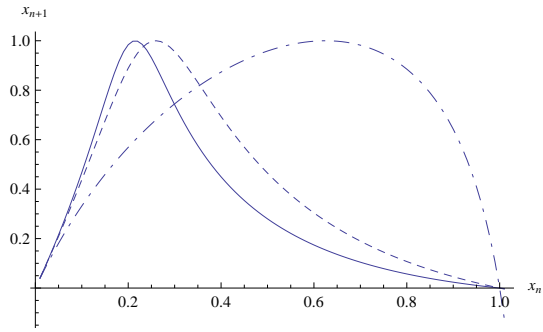

FIG. 1: Plots of $x_{n+1}$ with respect to $x_n$ for the q-logistic map, with $A = 4$ and $q = 0.05$, $0.1$ and $2.85$ for the bold, dashed and dot-dashed curves, respectively.

## II.　Q-DEFORMED LOGISTIC MAP

We propose a q-deformed logistic map by the non-linear difference equation

$$[x_{n+1}] = A[x_n](1 - [x_n]), \qquad (3)$$

where

$$[x] = \frac{1 - q^x}{1 - q}. \qquad (4)$$

Here $q$ is real and $x$ is in the interval $[0, 1]$. This q-deformed logistic map is different from Eq. (2). In the limit $q = 1$, it is seen that $[x] \to x$ and we obtain the usual logistic map (1). A partial motivation for our proposal (3) comes from the realization (see Introduction) that q-deformations effectively take into account the interactions. In the case of population growth, Eq. (1) relates to the population of a species assuming the mutual interactions among the members do not influence the number counted. This corresponds to an ideal case. However, in the growth or decay process, the mutual interactions might influence the number at an instant of time. This is effectively incorporated in (3). For instance, the difference between populations at unit interval of time, instead of $P_{n+1} = P_n + r_0 P_n (1 - \frac{P_n}{K})$, will be modified as $[P_{n+1}] = [P_n] + r_0 [P_n](1 - \frac{[P_n]}{K})$ which upon setting $[x_n] = \left(\frac{r_0}{1+r_0}\right)\frac{[P_n]}{K}$ gives (3) with $A = (1+r_0)$. Thus $[P_n] \neq P_n$ for $q \neq 1$ is the effective population. Another possibility is to consider $[x_{n+1}] = A[x_n][1 - x_n]$. This is symmetric when $x_n \to 1 - x_n$ and hence will not give concavity in parts of $x$- space as this is symmetric just as (1).

In fig. (1), we give the results of the q-logistic map (3) for various values of the deformation parameter $q$ and for $A = 4$. The bold and the dashed curves in fig. (1) clearly show that the q-deformed map, for small values of the deformation parameter $q$, is concave in parts of the $x$-space, whereas the usual logistic map ($q = 1$) is always convex. In the context of modelling population growth taking into account interactions between members, the q-logistic map suggests a scenario where the population growth is quick while the decay is much slower. This can be clearly seen from the bold curve in fig. (1) for a low value of $q$. This could be a reasonable model of the population evolution in a modern society, where with the advancement of medical technology, the population growth is much more rapid than its decay, in contrast to the usual logistic map where the two processes are roughly equal. For $q > 1$, concavity is not found, as can be seen from the dot-dashed curve in fig. (1).

## III.　FIXED POINTS OF THE Q-LOGISTIC MAP

From Eqs. (3) and (4), it follows

$$q^{x_{n+1}} = 1 - \frac{A(1 - q^{x_n})}{(1 - q)}(q^{x_n} - q). \qquad (5)$$

Let $q^{x_n} = X_n$ and $q^{x_{n+1}} = X_{n+1}$. This implies that $x_n = \frac{\ln X_n}{\ln q}$ and $x_{n+1} = \frac{\ln X_{n+1}}{\ln q}$, and $q \neq 1$. Then (5) becomes

$$X_{n+1} = 1 + A\frac{(X_n - 1)(X_n - q)}{(1 - q)}. \qquad (6)$$

(A). *One − cycle*:
In order to find the fixed points for the one-cycle, we set $x_{n+1} = x_n$ implying $X_{n+1} = X_n$. Using (6), this requirement becomes a quadratic equation for $X_n$

$$X_n(1-q) = 1 - q + A\{X_n^2 - X_n(1+q) + q\}, \tag{7}$$

and solving for $X_n$ we find

$$X_n = \frac{1}{2A}\{A(1+q) + 1 - q \pm \sqrt{(A(1+q) + 1 - q)^2 - 4A^2q - 4A + 4Aq}\,\}. \tag{8}$$

For real $q$, $X_n$ is real and so the expression under the square-root should be $\geq 0$. This expression actually simplifies to $(1-q)^2(A-1)^2$ which is $\geq 0$ for any real value of $A$. Thus the two roots of Eq. (7) are: $X_n = 1$ and $\frac{1}{A}(Aq+1-q)$. These roots correspond to (see below (5)) $q^{x_n} = 1$ and $q + \frac{1-q}{A}$. Consider the fixed point corresponding to $q^{x_n} = 1$. Clearly $x_n = 0$ is a fixed point. This is the *trivial fixed point* and is the same as in the usual logistic map (1). The other fixed point corresponding to $q^{x_n} = q + \frac{1-q}{A}$ is *non-trivial* and is given by

$$x_n = \frac{\ln(q + \frac{1-q}{A})}{\ln q}. \tag{9}$$

Thus the two fixed points of the one-cycle q-logistic map (3) are $x_n = 0$ and $x_n = \frac{\ln(q+\frac{1-q}{A})}{\ln q}$. In the limit $q \to 1$, as the q-logistic map (3) becomes the usual logistic map (1), the second fixed point becomes $x_n \to \frac{A-1}{A}$, reproducing the non-trivial fixed point of the usual logistic map (1) for one-cycle. Since $x_n$ is real, we get the condition $q + \frac{1-q}{A} > 0$. The non-trivial fixed point (9) is qualitatively and quantitatively different from that in the usual logistic map.

(B). *Two − cycle*:
In order to get the fixed points for the two-cycle for the q-logistic map (3), we set $x_{n+2} = x_n$ in (3) which implies $X_{n+2} = X_n$. Using (6), this requirement gives for $X_n$ the equation ($q \neq 1$)

$$(1 - X_n)(X_n - \frac{Aq + 1 - q}{A})\{A^2X_n^2 - AX_n(A + Aq - 1 + q) + A^2q - Aq(1-q) + (1-q)^2\} = 0. \tag{10}$$

The two fixed points of the one-cycle are given by the first two parenthetical terms in (10). The *new* fixed points are obtained by equating the term in the curly bracket, in the above equation, to zero. They are given by

$$X_n = \frac{1}{2A}\left(A(1+q) - 1 + q \pm (1-q)\sqrt{A^2 - 2A - 3}\right). \tag{11}$$

Since $x_n$ is real, $X_n$ is real. The expression $A^2 - 2A - 3$ is negative for $A < 3$. Thus to realize the new fixed points (bifurcation) $A$ must be greater than 3. This feature occurs in the usual logistic map (1) and therefore the fact that bifurcation in the logistic map begins after $A = 3$ appears to be universal. Thus the fixed points of the two-cycle are given by

$$\begin{aligned} X_n &= 1, \\ X_n &= \frac{Aq + 1 - q}{A}, \\ X_n &= \frac{1}{2A}\left(A(1+q) - 1 + q \pm (1-q)\sqrt{A^2 - 2A - 3}\right). \end{aligned} \tag{12}$$

At $A = 3$, the new fixed points coincide with $\frac{1+2q}{3}$ which is also the non-trivial one-cycle fixed point.

In fig. (2) is plotted the fixed points of the two-cycle normal logistic map with respect to the parameter $A$ while figs. (3) depict the fixed points of the two-cycle q-logistic map for different values of deformation parameter $q$. It can be seen from figs. (3) that for low values of the parameter $q$, the lower branch of the bifurcation diagram is tending, asymptotically, towards the fixed point 0, thereby exhibiting a tendency for the co-existence of attractors, though for the parameters shown it does not actually reach zero.

## IV. STABILITY ANALYSIS: LYAPUNOV EXPONENT

The q-logistic map (3) is written in terms of a function

$$\begin{aligned} f(x) &= A[x](1 - [x]) \\ &= \frac{A(1-q^x)}{(1-q)^2}(q^x - q), \end{aligned} \tag{13}$$

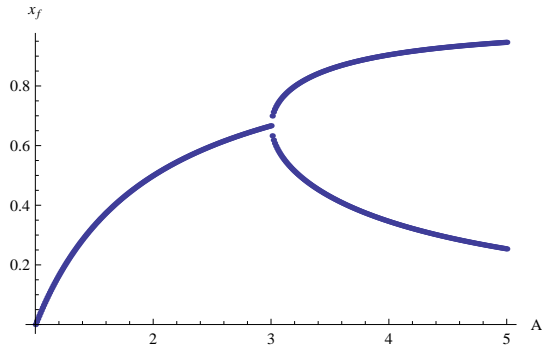

FIG. 2: Fixed points ($x_f$) of the two-cycle normal logistic map ($q = 1$) with respect to the parameter $A$.

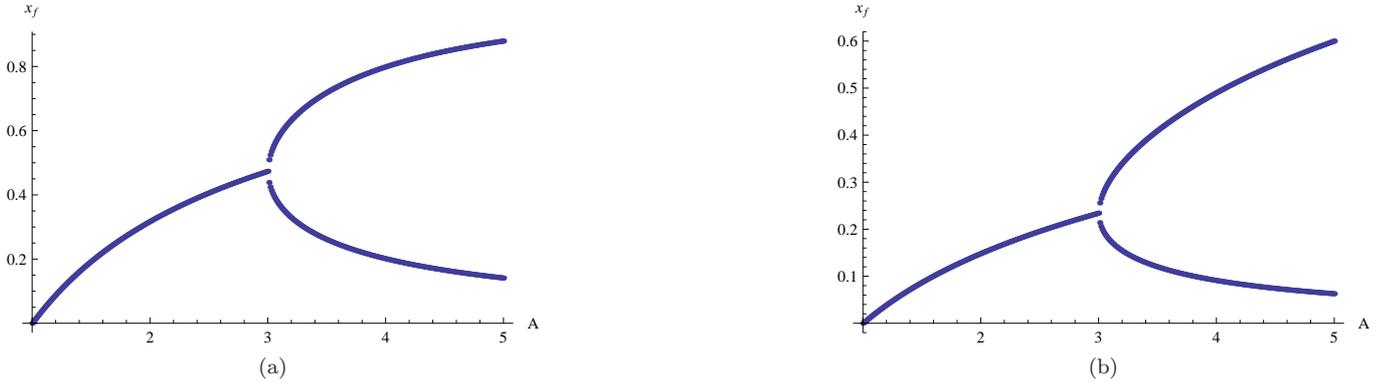

(a)  (b)

FIG. 3: Plots of the fixed points ($x_f$) of the two-cycle q-deformed logistic map with respect to the parameter $A$. Fig. (A) corresponds to $q = 0.2$, while fig. (B) corresponds to $q = 0.01$.

from which

$$f'(x) = \frac{A \ln q}{(1-q)^2} q^x (1 + q - 2q^x). \qquad (14)$$

Let us start with $x_0$. Then $f'(x_0) = \frac{A \ln q}{(1-q)^2} q^{x_0} (1 + q - 2q^{x_0})$. $x_1$ is computed using (3) as $[x_1] = A[x_0](1 - [x_0])$ which implies that $q^{x_1} = 1 - [x_1](1 - q)$. Then $f'(x_1) = \frac{A \ln q}{(1-q)^2} q^{x_1} (1 + q - 2q^{x_1})$. This process is continued to get $f'(x_2)$, $f'(x_3)$, …. The Lyapunov exponent is [22]

$$L = \left\{ \frac{1}{N} \sum_{k=0}^{N-1} \ln |f'(x_k)| \right\}_{N \to \infty}. \qquad (15)$$

In figs. (4), we show the Lyapunov exponent for the q-deformed logistic map with respect to the deformation parameter $q$ and parameter $A = 3.6$, i.e., in the regime of the two-cycle fixed point which occurs for $A > 3$. The figures bring out the advantage of having the parameter $q$, as with a change in the value of $q$ one is able to go from the chaotic to the regular dynamical regime. The figs. (6), where Lyapunov exponent is plotted with respect to the parameter $A$ for given values of $q$, complement the figs. (4) in that after $A > 3.6$ a predominantly chaotic behavior is seen. In fig. (5), the Lyapunov exponent is plotted with respect to $q$ and $A = 2.8$, i.e., in the regime of the one-cycle fixed point. A comparison of fig. (5) with (4(b)) shows that all other parameters, including the initial choice $x_0$, being same, a difference in $A$ from 3.6 (fig. (4(b))), i.e., in the two-cycle fixed point regime to $A = 2.8$, i.e., in the one-cycle fixed point regime, changes the dynamics from being predominantly chaotic (fig. (4(b))) to regular (fig. (5)).

## V. PARRONDO'S PARADOX: WEB DIAGRAM

Parrondo's paradox [26–29] is based on the combination of two negatively biased (losing) games which when combined give rise to a positively biased (winning) game and was originally used as an interpretation of the Brownian ratchet model [30, 31] in game-theoretic terms. These games have since found applications in many fields such as quantum information [32–34], stochastic resonance [35], random walks and diffusions [36] and quasi birth-death processes [37], among others.

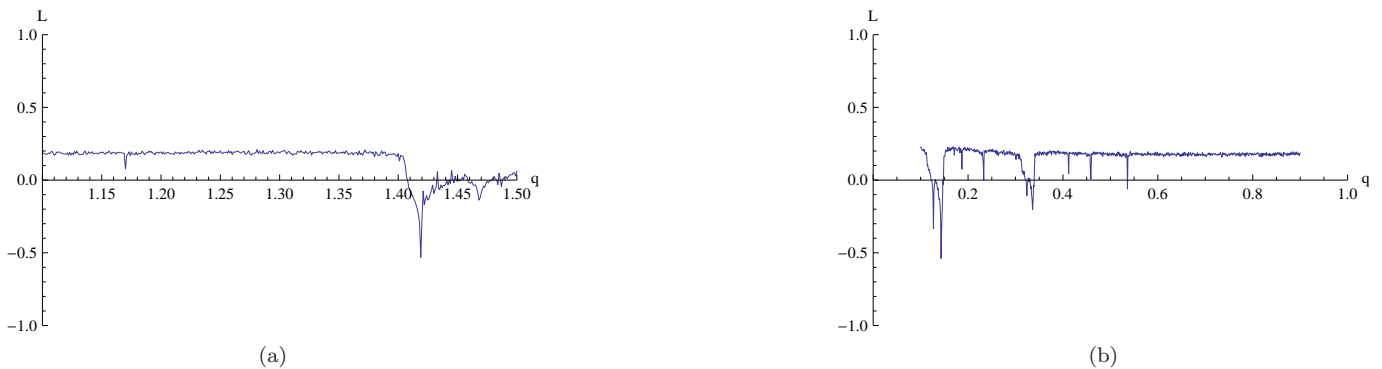

FIG. 4: Plots of Lyapunov exponent $L$ (15) with respect to the deformation parameter $q$. Here the parameter $A = 3.6$ and the initial choice $x_0 = 0.1$. Fig. (a) corresponds to $q$ from 1.1 to 1.5, while in fig. (b) the range of $q$ is from 0 to 0.99.

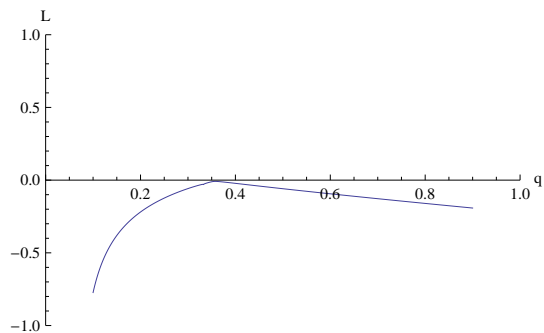

FIG. 5: Plot of Lyapunov exponent $L$ (15) with respect to the deformation parameter $q$. Here the parameter $A = 2.8$ and the initial choice $x_0 = 0.1$.

In the context of the q-logistic map, as seen from figs. (1), for smaller values of the deformation parameter $q$ there is a marked concavity in the curves which could be thought of as a quick growth accompanied by a slower decay. Thus the growth side could be considered as the winning strategy while the decay would be the loosing one. We use this analogy to illustrate Parrondo's paradox in the logistic map. This is studied here using the web diagram [38] for the map. A web diagram, also called a cobweb plot, is a graph that can be used to visualize successive iterations of a function $y = f(x)$. In particular, the segments of the diagram connect the points $(x, f(x))$, $(f(x), f(f(x)))$, $(f(f(x)), f(f(f(x))))$, .... The diagram gets its name from the fact that its straight line segments anchored to the functions $y = x$ and $y = f(x)$ can resemble a spider web.

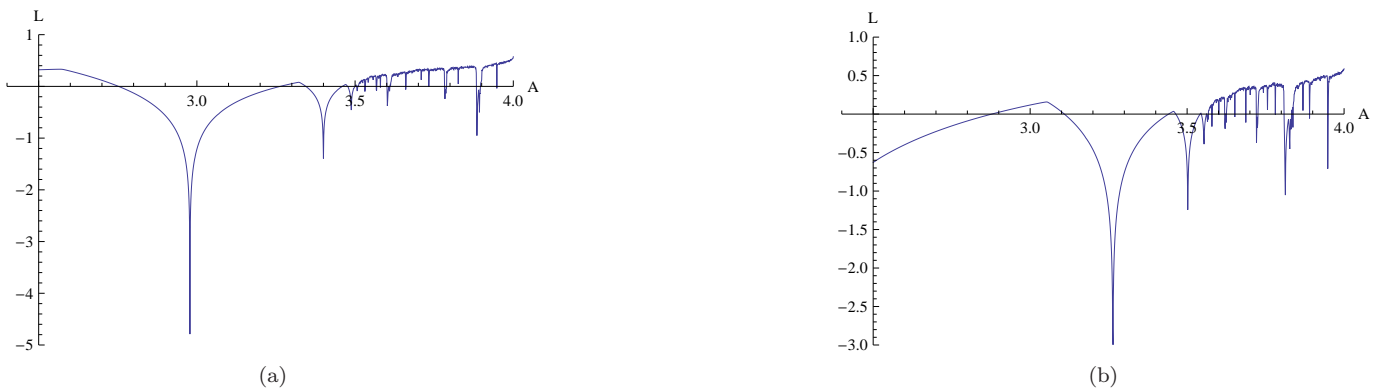

FIG. 6: Plots of Lyapunov exponent $L$ (15) with respect to the parameter $A$ (3). Here the deformation parameter $q = 0.15$ and 1.40 for the figs. (a) and (b), respectively. The initial choice $x_0 = 0.1$.

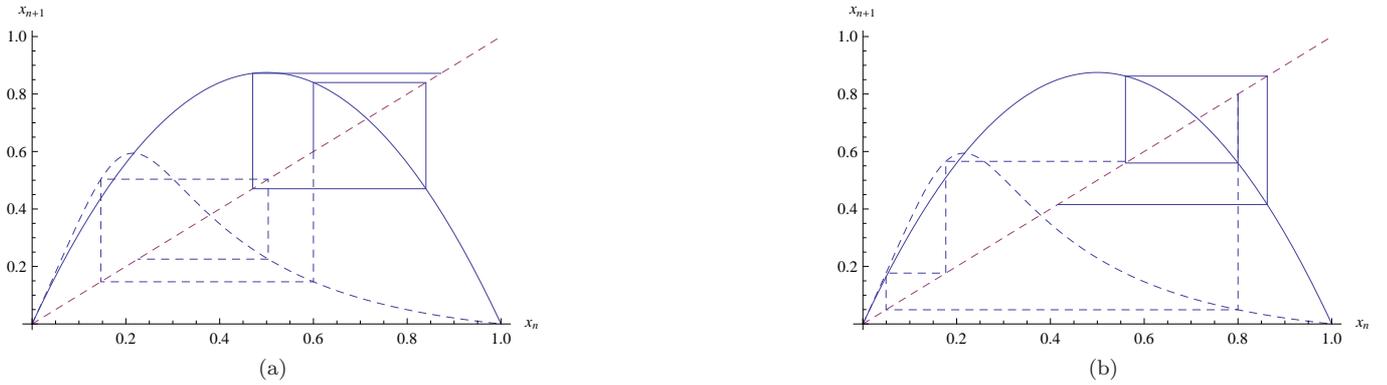

FIG. 7: Web diagram of the logistic map with the parameter $A = 3.5$, iterations equal to 3 and initial choice $x = 0.6$ and $0.8$ for figs. (a) and (b), respectively. In each figure, the solid curve corresponds the usual logistic map ($q = 1$), while the dashed curve corresponds to the q-logistic map with $q = 0.05$.

To illustrate the Parrondo paradox, we start in the fig. (7(a)) and (b) from the loosing (decaying) side, i.e., we choose the initial value of $x$ on the $X$-axis such that the corresponding logistic function falls in the decaying side. With the help of the web diagrams, it can be easily seen that starting from the losing side, one is able to end up in the winning side in a fewer number of iterations for the q-deformed map as compared to the normal map. Thus, for e.g., a comparison of the dashed (q-deformed map) and the solid (usual map) in fig. (7(a)) clearly shows that while in the q-logistic map one is able to, starting from the loosing side with initial choice $x = 0.6$, reach the winning side in two iterations of the web diagram, it takes the usual map three iterations to reach a point which is just at the boundary of the winning and loosing side. This is made more dramatic in fig. (7(b)), where everything else remaining the same as in fig. (a), the initial choice is moved more into the losing side, i.e., $x = 0.8$. This, paradoxically, improves the winning chances for the q-deformed map in that after two iterations, the web diagram reaches into the early rising part of the curve, i.e., a very comfortable winning side. In contrast, the normal logistic map has a deterioration in its performance in that even after three iterations the web diagram is not able to reach the winning side. This could be envisaged as a result of the concavity introduced in the q-deformed map for low values of $q$, as seen from the bold curve in fig. (1). This actually aids in the improvement of performance with an initial choice more into the losing side. The usual, undeformed, logistic map does not have this feature and hence with the initial choice more into the losing side, the chance of getting into the winning side is also depleted. This thus puts in perspective the fact that though Parrondo's paradox can be exhibited by the usual logistic map, its effect is much more pronounced for the case of the q-deformed map, a feature which could have many interesting implications.

## VI. SUMMARY

We have proposed a q-deformed logistic map by the non-linear difference equation (3) and found *concavtiy* in parts of the $x$-space for small values of the deformation parameter $q$, whereas the usual logistic map is always convex. Concavity is not seen for $q > 1$. For $q < 1$, the proposed q-deformed logistic map (3) could serve as a reasonable model of the evolution of population in a modern society. The fixed points of the proposed q-logistic map for one and two cycles are obtained. The non-trivial fixed points are seen to be qualitatively and quantitatively different from those of the usual logistic map. In the case of two cycles, the bifurcation happens after $A = 3$ as in the case of the usual logistic map. Thus the value of $A = 3$ appears to be *universal*. For $q \ll 1$, we find a tendency for the co-existence of attractors. The stability of the proposed map is studied using Lyapunov exponent. It is seen that with a change in the value of the deformation parameter $q$, a transition from the chaotic to the regular dynamical regime could be effected. For $q = 0.15$ and $1.4$, for $A < 3$, the map is Lyapunov stable and becomes chaotic for $A > 3$, consistent with the beginning of bifurcation for $A$ above 3. We have also attempted to understand Parrondo's paradox using the logistic map, making use of a web diagram for the map. By starting from the "loosing" side, the q-logistic map allows the passage to the "winning" side by using fewer iterations than the usual logistic map. This feature is especially brought out by considering the case where by starting from a position more into the losing side, one is able to reach the winning side, in the q-logistic map, in a few iterations whereas starting from the same initial condition, in the normal logistic map, the winning side is not reached even in a larger number of iterations of the map. We expect this feature to have wider implications.

## Acknowledgement

We thank R. Jagannathan, V. V. Sreedhar for useful discussions. We would also like to thank Awadhesh Prasad for going through the manuscript and making useful suggestions.

---